\documentclass[a4paper,12pt]{article}
\usepackage{a4,amsmath,amssymb,axodraw,slashed,epsfig}
\usepackage{axodraw}

\begin{document}

\vspace*{-1.5cm}

\begin{flushright}
PSI--PR--16--10
\end{flushright}
\begin{center}
{\large \bf Effective Multi--Higgs Couplings to Gluons} \\[0.3cm]

Michael Spira \\[0.3cm]

{\it Paul Scherrer Institut, CH--5232 Villigen PSI, Switzerland}
%\\[0.5cm]
\end{center}

\begin{abstract}
\noindent
Standard-Model Higgs bosons are dominantly produced via the gluon-fusion
mechanism $gg \to H$ at the LHC, i.e.~in a loop-mediated process with
top loops providing the dominant contribution. For the measured Higgs
boson mass of $\sim 125$ GeV the limit of heavy top quarks provides a
reliable approximation as long as the relative QCD corrections are
scaled with the full mass-dependent LO cross section. In this limit
the Higgs coupling to gluons can be described by an effective
Lagrangian.  The same approach can also be applied to the coupling of
more than one Higgs boson to gluons. We will derive the effective
Lagrangian for multi-Higgs couplings to gluons up to N$^4$LO thus
extending previous results for more than one Higgs boson. Moreover we
discuss gluonic Higgs couplings up to NNLO, if several heavy quarks
contribute.
\end{abstract}

\section{Introduction}
%        ============
The discovery of a resonance with 125 GeV mass \cite{discovery} that is
compatible with the Standard-Model (SM) Higgs boson \cite{couplings}
marked a milestone in particle physics.  The existence of the Higgs
boson is inherently related to the mechanism of spontaneous symmetry
breaking \cite{hi64} while preserving the full gauge symmetry and the
renormalizability of the SM \cite{smren}. The dominant production
process of the Higgs boson at the LHC is the loop-induced gluon-fusion
process mediated by top-quark loops and to a lesser extent bottom- and
charm-quark loops \cite{gghlo}.  The QCD corrections are known up to
N$^3$LO in the limit of heavy top quarks \cite{gghnlolim, gghnnlo,
gghn3lo}, while the full quark mass dependence is only known up to NLO
\cite{gghnlo,gghnlo0}. At NNLO subleading terms in the large top mass
expansion \cite{gghnnlomt} and leading contributions to the top+bottom
interference \cite{tbint} are known. The limit of heavy top quarks has
also been adopted for threshold-resummed calculations
\cite{limit,gghresum}, while the inclusion of finite quark-mass effects
in the resummation has been considered recently \cite{gghresummt}. It
has been shown that the limit of heavy top quarks $m_t^2 \gg M_H^2$
provides a reasonable approximation to the calculation of the
gluon-fusion cross section with full mass dependence as long as the
relative QCD corrections are scaled with the fully massive LO cross
section \cite{gghnlo,limit}. In the heavy-top-quark limit the
calculation of the gluon-fusion cross section can be simplified by
starting from an effective Lagrangian describing the Higgs coupling to
gluons after integrating out the top contribution \cite{let}. The same
approach has also been applied to Higgs pair production via gluon
fusion, $gg \to HH$, at NLO \cite{gghhnlo}, NNLO
\cite{gghhnnlo1,gghhnnlo2} as well as to threshold resummation up to
NNLL \cite{gghhnnll}. It has been shown that finite mass effects amount
to about 5\% in the single Higgs case and $15\%$ for Higgs boson pairs
\cite{gghhnlomt,gghhnlomt0}.

In this letter we will derive the effective Lagrangian for multi-Higgs
couplings to gluons to N$^4$LO for arbitrary numbers of external Higgs
bosons thus extending previous work beyond the single-Higgs case.  In
Section 2 we will discuss and present the effective Lagrangian for the
SM Higgs boson up to N$^4$LO, while Section 3 will extend this analysis
to an arbitrary number of heavy quarks contributing to the gluonic Higgs
coupling up to NNLO. In Section 4 we will conclude.

\section{Standard-Model Higgs Bosons}
%        ============================
The starting point for the derivation of the effective Lagrangian in the
heavy-top-quark limit is the low-energy limit of the top-quark
contributions to the Wilson coefficient of the gluonic field-strength
operator $\hat G^{a\mu\nu} \hat G^a_{\mu\nu}$, where $\hat G^{a\mu\nu}$
denotes the ($\overline{\rm MS}$-subtracted) gluonic operator of
colour-SU(3) in the low-energy limit with 6 active flavours\footnote{The
same ansatz has also been used in the derivation of the effective $Hgg$
coupling in Refs.~\cite{limit}.},
\begin{equation}
{\cal L}_{g} = -\frac{1-\Pi_t}{4} \hat G^{a\mu\nu} \hat G^a_{\mu\nu}
\label{eq:leffgg}
\end{equation}
The Wilson coefficient $\Pi_t$ denotes the gauge-invariant vacuum
polarization function of the gluon that is determined by the top-quark
contribution to the gluon self-energy and the two-point-function parts
of the external vertices attached to the gluons. This boils down to the
inverse top-quark contribution to the strong coupling constant so that
$\Pi_t$ is related to the decoupling relation between the strong
coupling constant in an $(N_F+1)$- and $N_F$-flavour theory ($N_F=5$),
\begin{equation}
\alpha_s^{(N_F)}(\mu_R^2) =
\zeta_{\alpha_s}~\alpha_s^{(N_F+1)}(\mu_R^2) \, , \qquad \qquad
\zeta_{\alpha_s} = 1 + \sum_n D_n
\left(\frac{\alpha_s^{(N_F+1)}(\mu_R^2)}{\pi}\right)^n
\label{eq:decoup}
\end{equation}
with the perturbative coefficients up to fourth order \cite{decoup0,
decoup1, decoup} [$L_t = \log(\mu_R^2/\overline{m_t}^2(\mu_R^2))$]
\begin{eqnarray}
D_1 & = & -\frac{1}{6} L_t \qquad \qquad \qquad \qquad \qquad \qquad
D_2 = \frac{11}{72} - \frac{11}{24} L_t + \frac{1}{36} L_t^2 \\
D_3 & = & \frac{564731}{124416} - \frac{82043}{27648} \zeta_3 -
\frac{2633}{31104} N_F - \frac{955-67 N_F}{576} L_t + \frac{53-16
N_F}{576} L_t^2 - \frac{1}{216} L_t^3 \nonumber \\
D_4 & = & \frac{291716893}{6123600} - \frac{121}{4320} \log^5 2 +
\frac{3031309}{1306368} \log^4 2 + \frac{121}{432} \zeta_2 \log^3 2 -
\frac{3031309}{217728} \zeta_2 \log^2 2 \nonumber \\
& + & \frac{2057}{576} \zeta_4 \log 2  + \frac{1389}{256} \zeta_5 -
\frac{76940219}{2177280} \zeta_4 - \frac{2362581983}{87091200} \zeta_3 +
\frac{3031309}{54432} a_4 + \frac{121}{36} a_5 \nonumber \\
& - & \frac{151369}{2177280} X_0 + N_F \left(-\frac{4770941}{2239488} +
\frac{685}{124416} \log^4 2 - \frac{685}{20736} \zeta_2 \log^2 2 +
\frac{3645913}{995328} \zeta_3 \right. \nonumber \\
& & \left. - \frac{541549}{165888} \zeta_4 +
\frac{115}{576} \zeta_5 + \frac{685}{5184} a_4\right) + N_F^2
\left(-\frac{271883}{4478976} + \frac{167}{5184} \zeta_3\right)
\nonumber \\
& -& \left[\frac{7391699}{746496} + \frac{2529743}{165888} \zeta_3 + N_F
\left( \frac{110341}{373248} - \frac{110779}{82944} \zeta_3\right) -
N_F^2 \frac{6865}{186624} \right] L_t \nonumber \\
& + & \left( \frac{2177}{3456} - N_F \frac{1483}{10368} - N_F^2
\frac{77}{20736} \right) L_t^2 -\left(\frac{1883}{10368} + N_F
\frac{127}{5184} - \frac{N_F^2}{324} \right) L_t^3 + \frac{L_t^4}{1296}
\nonumber
\end{eqnarray}
where $\overline{m_t}^2(\mu_R^2))$ denotes the $\overline{\rm MS}$ top
mass at the renormalization scale $\mu_R$.  The constants used in this
expression are given by $a_n = Li_n(1/2)$ and $X_0 =
1.8088795462...$ . The decoupling coefficient contains
one-par\-ti\-cle-reducible contributions and the Wilson coefficient
of the Lagrangian Eq.~(\ref{eq:leffgg}) is obtained from the
inverse,
\begin{equation}
\Pi_t = 1 - \frac{1}{\zeta_{\alpha_s}} = \sum_n C_n
\left(\frac{\alpha_s^{(N_F+1)}}{\pi}\right)^n
\label{eq:pit}
\nopagebreak
\end{equation}
with the perturbative coefficients up to fifth order
\begin{eqnarray}
C_1 & = & -\frac{1}{6} L_t \qquad \qquad \qquad \qquad \qquad \qquad
C_2 = \frac{11}{72} - \frac{11}{24} L_t \nonumber \\
C_3 & = & \frac{564731}{124416} - \frac{82043}{27648} \zeta_3 -
\frac{2633}{31104} N_F - \frac{2777-201 N_F}{1728} L_t - \frac{35+16
N_F}{576} L_t^2 \nonumber \\
C_4 & = & \frac{1166295847}{24494400} - \frac{121}{4320} \log^5 2 +
\frac{3031309}{1306368} \log^4 2 + \frac{121}{432} \zeta_2 \log^3 2 -
\frac{3031309}{217728} \zeta_2 \log^2 2 \nonumber \\
& + & \frac{2057}{576} \zeta_4 \log 2  + \frac{1389}{256} \zeta_5 -
\frac{76940219}{2177280} \zeta_4 - \frac{2362581983}{87091200} \zeta_3 +
\frac{3031309}{54432} a_4 + \frac{121}{36} a_5 \nonumber \\
& - & \frac{151369}{2177280} X_0 + N_F \left(-\frac{4770941}{2239488} +
\frac{685}{124416} \log^4 2 - \frac{685}{20736} \zeta_2 \log^2 2 +
\frac{3645913}{995328} \zeta_3 \right. \nonumber \\
& & \left. - \frac{541549}{165888} \zeta_4 +
\frac{115}{576} \zeta_5 + \frac{685}{5184} a_4\right) + N_F^2
\left(-\frac{271883}{4478976} + \frac{167}{5184} \zeta_3\right)
\nonumber \\
& + & \left[\frac{2875235}{248832} - \frac{897943}{55296} \zeta_3 - N_F
\left( \frac{40291}{124416} - \frac{110779}{82944} \zeta_3\right) +
N_F^2 \frac{6865}{186624} \right] L_t \nonumber \\
& - & \left( \frac{1333}{10368} + N_F \frac{1081}{10368} + N_F^2
\frac{77}{20736} \right) L_t^2
-\left(\frac{1697}{10368} + N_F \frac{175}{5184} - N_F^2 \frac{1}{324}
\right) L_t^3 \nonumber \\
C_5 & = & C_{50} + \left(-\frac{685}{10368}
N_F^{2} a_4 -\frac{11679301}{ 435456} N_F a_4+\frac{93970579}{217728}
a_4-\frac{121}{72} N_F a_5 +\frac{3751}{144} a_5 \right. \nonumber \\
& & \left. +\frac{121 }{8640} N_F \log^5 2 -\frac{3751}{17280} \log^5
2-\frac{685}{248832} N_F^{2} \log^4 2 -\frac{11679301}{ 10450944} N_F \log^4
2 \right. \nonumber \\
& & \left. +\frac{93970579}{5225472} \log^4 2-\frac{121}{864} N_F
\zeta_2 \log^3 2 +\frac{3751}{1728} \zeta_2 \log^3 2 +\frac{685}{41472}
N_F^{2} \zeta_2 \log^{2}2 \right. \nonumber \\
& & \left. +\frac{11679301}{1741824} N_F \zeta_2 \log^{2}2
-\frac{93970579}{870912} \zeta_2 \log^{2}2 -\frac{2057}{1152} N_F
\zeta_4 \log 2 +\frac{63767}{2304} \zeta_4 \log 2 \right. \nonumber \\
& & \left. -\frac{211}{10368} N_F^{3} \zeta_3+\frac{270407}{8957952}
N_F^{3}-\frac{4091305}{1990656} N_F^{2} \zeta_3+\frac{576757}{331776}
N_F^{2} \zeta_4+\frac{115}{2304} N_F ^{2} \zeta_5 \right. \nonumber \\
& & \left. +\frac{48073}{165888} N_F^{2}+\frac{151369}{4354560} N_F X_0+
\frac{12171659669}{232243200} N_F \zeta_3-\frac{608462731}{69672960} N_F
\zeta_4 \right. \nonumber \\
& & \left. -\frac{313489}{41472} N_F
\zeta_5-\frac{75861299783}{3135283200} N_F-\frac{4692439 }{8709120}
X_0-\frac{4660543511}{19353600} \zeta_3 \right. \nonumber \\
& & \left. -\frac{4674213853}{17418240} \zeta_4+\frac{807193}{10368}
\zeta_5+\frac{846138861149}{3135283200}\right) L_t
+ \left(-\frac{481}{62208} N_F^{3}- \frac{28297}{110592}N_F^{2} \zeta_3
\right.  \nonumber \\
& & \left.
+\frac{373637}{746496} N_F^{2}+\frac{2985893}{331776} N_F
\zeta_3-\frac{47813}{4608} N_F
-\frac{26296585}{442368} \zeta_3+
\frac{143939741}{1990656}\right) L_t^{2} \nonumber \\
& + & \left(\frac{77}{124416} N_F^{3}+\frac{175}{27648}
N_F^{2}-\frac{5855}{ 124416} N_F-\frac{130201}{124416}\right) L_t^{3}
\nonumber \\
& + & \left(-\frac{1}{2592} N_F^{3}+\frac{47}{4608}
N_F^{2}-\frac{317}{6912} N_F-\frac{51383}{ 165888}\right) L_t^{4}
\end{eqnarray}
where the logarithms of the coefficient $C_5$ have been reconstructed
from the result of Ref.~\cite{decoup} including the recent five-loop
result of the QCD beta function \cite{beta5} (partly confirmed by
\cite{beta5p}). The constant $C_{50}$ is irrelevant for our derivation
of the effective Lagrangian for gluonic Higgs couplings. Note that the
highest powers of the logarithmic $L_t$ terms disappeared in this
expression as required by the proper RG-evolution of the
one-particle-irreducible part $\Pi_t$.  Using the low-energy theorem for
a light Higgs boson \cite{let} the effective top-quark contribution to
the Lagrangian of Eq.~(\ref{eq:leffgg}) is related to the couplings of
external Higgs bosons in the heavy-top-quark limit by the
replacement\footnote{In the case of an extended Higgs sector with
several scalar Higgs bosons coupling to the top quark the replacement
$\overline{m_t}(\mu_R^2) \to \overline{m_t}(\mu_R^2) (1+\sum_i c_i
H_i/v)$ has to be implemented, where $c_i$ are the top quark Yukawa
couplings normalized to the SM coupling. This results in the
correspondence $H/v \leftrightarrow \sum_i c_i H_i/v$ for all subsequent
steps.} $\overline{m_t}(\mu_R^2) \to \overline{m_t}(\mu_R^2) (1+H/v)$,
i.e.
\begin{equation}
L_t \to \bar L_t = L_t - 2\log\left(1 + \frac{H}{v} \right) \qquad
\mbox{and} \qquad
\Pi_t \to \bar \Pi_t
\label{eq:shift}
\end{equation}
where $H$ denotes the physical Higgs field, $v$ the vacuum expectation
value and $\bar \Pi_t$ the contribution to the Wilson coefficient with
the shifted top-quark mass\footnote{Note that diagrammatically for the
single-Higgs case this expression coincides with the replacement
$\frac{1}{\not\!\;p - m_t} \to \frac{1}{\not\!\;p - m_t} \frac{m_t}{v}
\frac{1}{\not\!\;p - m_t}$ of the top-quark propagators inside the
gluonic correlation functions up to 4th order in the gluon fields at the
point where $m_t$ is either the unrenormalized or the pure
$\overline{\rm MS}$ mass \cite{gghnlo}.}.  Based on this replacement it
is obvious that only the logarithmic $L_t$ terms of $\Pi_t$ are relevant
for the effective gluonic Higgs couplings.  The object $\bar \Pi_t$ is
expressed in terms of the $(N_F+1)$-flavour coupling
$\alpha_s^{(N_F+1)}$. To derive the low-energy Lagrangian in the
$N_F$-flavour theory we have to transform the $(N_F+1)$-flavour coupling
into the $N_F$-flavour one by means of the relation \cite{decoup0,
decoup1, decoup}
\begin{eqnarray}
\alpha_s^{(N_F+1)}(\mu_R^2) & = & \alpha_s^{(N_F)}(\mu_R^2) \left\{ 1 +
\frac{\alpha_s^{(N_F)}(\mu_R^2)}{\pi} \frac{L_t}{6} + \left(
\frac{\alpha_s^{(N_F)}(\mu_R^2)}{\pi} \right)^2 \left[ -\frac{11}{72} +
\frac{11}{24} L_t + \frac{L_t^2}{36} \right] \right. \nonumber \\
& + & \left. \left( \frac{\alpha_s^{(N_F)}(\mu_R^2)}{\pi} \right)^3
\left[ -\frac{564731}{124416} + \frac{82043}{27648} \zeta_3+
\frac{2633}{31104} N_F \right. \right. \nonumber \\
& & \left. \left. +\left( \frac{2645}{1728}- \frac{67}{576} N_F \right)
L_t +\left(\frac{167}{576}+\frac{N_F}{36}\right) L_t^2
+\frac{L_t^3}{216} \right]
+ {\cal O}(\alpha_s^4) \right\}
\end{eqnarray}
derived from inverting Eq.~(\ref{eq:decoup}). For the proper low-energy
limit the gluonic field-strength operator is expressed in terms of the
one with $N_F=5$ active flavours which leads to a global factor
$\zeta_{\alpha_s}$ so that the kinetic term of the gluons is properly
normalized in the low-energy limit\footnote{Diagrammatically this step
corresponds to adding the external $\overline{\rm MS}$-renormalized
self-energies and two-point-function contributions to the vertices
involving top quarks at vanishing external momentum.}. In this way we
arrive at the low-energy Lagrangian in terms of the top $\overline{\rm
MS}$ mass.  The effective N$^4$LO Lagrangian for (multi-)Higgs couplings
to gluons reads finally
\begin{eqnarray}
{\cal L}_{eff} & = & \frac{\alpha_s}{12\pi} \left\{ (1 + \delta) \log
\left(1+\frac{H}{v}\right) - \frac{\eta}{2} \log^2
\left(1+\frac{H}{v}\right) \right. \nonumber \\
& & \hspace*{2cm} \left. + \frac{\rho}{3} \log^3 \left(1+\frac{H}{v}\right)
- \frac{\sigma}{4} \log^4\left(1+\frac{H}{v}\right)
\right\} G^{a\mu\nu} G^a_{\mu\nu}
\label{eq:leff}
\end{eqnarray}
with the QCD corrections up to N$^4$LO
\begin{eqnarray}
\delta & = & \delta_1 \frac{\alpha_s}{\pi} + \delta_2 \left(
\frac{\alpha_s}{\pi} \right)^2 + \delta_3 \left(
\frac{\alpha_s}{\pi} \right)^3 + \delta_4 \left(
\frac{\alpha_s}{\pi} \right)^4 + {\cal O}(\alpha_s^5) \nonumber \\
\eta & = & \eta_2 \left( \frac{\alpha_s}{\pi} \right)^2 + \eta_3 \left(
\frac{\alpha_s}{\pi} \right)^3 + \eta_4 \left(
\frac{\alpha_s}{\pi} \right)^4 + {\cal O}(\alpha_s^5) \nonumber \\
\rho & = & \rho_3 \left( \frac{\alpha_s}{\pi} \right)^3 + \rho_4 \left(
\frac{\alpha_s}{\pi} \right)^4 + {\cal O}(\alpha_s^5) \nonumber \\
\sigma & = & \sigma_4 \left( \frac{\alpha_s}{\pi} \right)^4
+ {\cal O}(\alpha_s^5)
\end{eqnarray}
The explicit perturbative coefficients are given by
\begin{eqnarray}
\delta_1 & = & \frac{11}{4} \qquad \qquad \qquad \qquad \qquad \qquad
\delta_2 = \frac{2777}{288} + \frac{19}{16} L_t + N_F
%\delta_1 & = & \frac{11}{4} \nonumber \\
%\delta_2 & = & \frac{2777}{288} + \frac{19}{16} L_t + N_F
\left(\frac{L_t}{3}-\frac{67}{96} \right) \nonumber \\
\delta_3 & = & \frac{897943}{9216} \zeta_3 - \frac{2892659}{41472} +
\frac{209}{64} L_t^2 + \frac{1733}{288} L_t \nonumber \\
& + & N_F \left(\frac{40291}{20736} - \frac{110779}{13824} \zeta_3 +
\frac{23}{32} L_t^2 + \frac{55}{54} L_t \right) + N_F^2
\left(-\frac{L_t^2}{18} + \frac{77}{1728} L_t - \frac{6865}{31104}
\right) \nonumber \\
\delta_4 & = & 
-\frac{121}{1440} N_F \log^{5} 2 +\frac{3751}{2880}
\log^{5} 2+\frac{685}{41472} N_F^2 \log^{4} 2 +\frac{11679301}{
1741824} N_F \log^{4} 2 \nonumber \\
& - & \frac{93970579}{870912}
\log^{4} 2+\frac{121}{144} N_F \zeta_2 \log^{3} 2-\frac{3751}{288}
\zeta_2 \log^{3} 2-\frac{685}{6912} N_F^{2} \zeta_2 \log^{2 } 2
\nonumber \\
& - & \frac{11679301}{290304} N_F \zeta_2 \log^{2} 2 +
\frac{93970579}{145152} \zeta_2 \log^{2} 2 +\frac{2057}{192} N_F \zeta_4
\log 2 - \frac{63767}{384} \zeta_4 \log 2 \nonumber \\
& + & \frac{685}{1728} N_F^{2} a_4
+\frac{11679301}{72576} N_F a_4 -\frac{93970579}{36288}
a_4+\frac{121}{12} N_F a_5 -\frac{3751}{24} a_5+\frac{211}{1728} N_F^{3}
\zeta_3 \nonumber \\
& - & \frac{270407}{1492992} N_F^{3}+\frac{4091305}{331776} N_F^{2}
\zeta_3-\frac{576757}{ 55296} N_F^{2} \zeta_4-\frac{115}{384} N_F^{2}
\zeta_5-\frac{48073}{27648} N_F^{2} \nonumber \\
& - & \frac{151369}{725760} N_F
X_0-\frac{12171659669}{38707200} N_F \zeta_3+\frac{608462731}{ 11612160}
N_F \zeta_4+\frac{313489}{6912} N_F
\zeta_5 \nonumber \\
& + & \frac{76094378783}{522547200} N_F+\frac{4692439}{1451520}
X_0+\frac{28121193841}{19353600} \zeta_3+\frac{4674213853}{ 2903040}
\zeta_4-\frac{807193}{1728} \zeta_5 \nonumber \\
& - & \frac{854201072999}{522547200}
+ \left(\frac{481}{5184}
N_F^{3}+\frac{28297}{9216} N_F^{2} \zeta_3-\frac{21139}{3456}
N_F^{2}-\frac{32257}{288} N_F \zeta_3 \right. \nonumber \\
& & \left. + \frac{5160073}{41472} N_F+
\frac{9364157}{12288}
\zeta_3-\frac{49187545}{55296}\right) L_t
+ \left(-\frac{77}{6912}
N_F^{3}-\frac{1267}{13824} N_F^{2}+\frac{4139}{2304}
N_F \right. \nonumber \\
& & \left. +\frac{8401}{384}\right) L_t^2
+ \left(\frac{1}{108} N_F^{3}-\frac{157}{576} N_F^{2}+\frac{275}{192}
N_F+\frac{2299}{256} \right) L_t^3
\end{eqnarray}
and
\begin{eqnarray}
\eta_2 & = & \frac{35}{24} + \frac{2}{3} N_F \nonumber \\
%\eta_3 & = & \frac{1333}{432} + \frac{1081}{432} N_F + \frac{77}{864}
%N_F^2 + \left( \frac{589}{48} + \frac{191}{72} N_F - \frac{2}{9} N_F^2
%\right) L_t \nonumber \\
\eta_3 & = & \frac{1333}{432} + \frac{589}{48} L_t + N_F \left(
\frac{1081}{432} + \frac{191}{72} L_t \right) + N_F^2 \left( \frac{77}{864}
- \frac{2}{9} L_t \right) \nonumber \\
\eta_4 & = & 
\frac{481}{2592} N_F^{3}+N_F^{2} \left( \frac{28297}{4608}
\zeta_3-\frac{373637}{31104} \right)
+N_F \left(\frac{429965}{1728}-\frac{2985893}{13824} \zeta_3 \right)
\nonumber \\
& + & \frac{26296585}{18432} \zeta_3-
\frac{143976701}{82944}
+ \left(-\frac{77}{1728} N_F^{3}-\frac{1421}{3456}
N_F^{2}+\frac{9073}{1728} N_F+\frac{45059}{576} \right) L_t \nonumber \\
& + & \left(\frac{N_F^{3}}{18} -\frac{455}{288} N_F^{2}+\frac{63}{8}
N_F+\frac{6479}{128}\right) L_t^2
\nonumber \\
\rho_3 & = & \frac{1697}{144} + \frac{175}{72} N_F - \frac{2}{9} N_F^2
\nonumber \\
\rho_4 & = & \frac{130201}{1728} + \frac{18259}{192} L_t + N_F\left(
\frac{5855}{1728} + \frac{2077}{144} L_t \right) - N_F^2\left(
\frac{175}{384} + \frac{439}{144} L_t \right) \nonumber \\
& + & N_F^3 \left( \frac{L_t}{9} - \frac{77}{1728} \right)
\nonumber \\
\sigma_4 & = & \frac{51383}{864} + \frac{317}{36} N_F - \frac{47}{24}
N_F^2 + \frac{2}{27}N_F^3
\end{eqnarray}
where $G^a_{\mu\nu}$ denotes the gluon field strength tensor and
$\alpha_s$ the strong coupling constant with $N_F=5$ active flavours.
Note that in accordance with the RG-evolution the coefficients
$\delta_1, \eta_2, \rho_3$ and $\sigma_4$ are free of $L_t$ terms.
Numerically we obtain for $N_F=5$ light flavours
\begin{eqnarray}
\delta_1 & = & 2.75 \qquad \qquad \qquad \qquad \qquad \qquad \qquad
\qquad \qquad \quad
\delta_2 = 6.1528 + 2.8542 L_t \nonumber \\
\delta_3 & = & 3.4043 + 12.2240 L_t + 5.4705 L_t^2 \nonumber \\
\delta_4 & = & 36.0373 - 73.5997 L_t + 27.1760 L_t^2 + 10.4851 L_t^3
\nonumber \\
\eta_2 & = & 4.7917 \qquad \qquad \qquad \qquad \qquad \qquad \qquad
\qquad \qquad
\eta_3 = 17.8252 + 19.9792 L_t \nonumber \\
\eta_4 & = & -167.5239 + 88.6311 L_t + 57.4401 L_t^2 \qquad \qquad \quad
\rho_3 = 18.3819 \nonumber \\
\rho_4 & = & 75.3261 + 104.8906 L_t \qquad \qquad \qquad \qquad \qquad
\quad \, \,
\sigma_4 = 63.7998
\end{eqnarray}
If the running $\overline{\rm MS}$ top mass is replaced by the top pole
mass $M_t$ \cite{msbarpole}\footnote{Note that the low-energy strong
coupling constant with $N_F=5$ active flavours is used in this
relation.} [i.e.~$L_t = \log (\mu_R^2/M_t^2)$ is used everywhere],
\begin{eqnarray}
\overline{m_t}(\mu_R^2) & = & M_t \left\{ 1 - \left(\frac{4}{3} + \log
\frac{\mu_R^2}{M_t^2} \right) \frac{\alpha_s^{(N_F)}(\mu_R^2)}{\pi}  + \left[
-\frac{3019}{288} - 2\zeta_2 - \frac{2}{3} \zeta_2 \log 2 +
\frac{\zeta_3}{6} \right. \right. \nonumber \\
& & \left. \left. - \frac{461}{72} \log \frac{\mu_R^2}{M_t^2} -
\frac{23}{24} \log^2 \frac{\mu_R^2}{M_t^2} + N_F \left( \frac{71}{144}
+ \frac{\zeta_2}{3} + \frac{13}{36} \log \frac{\mu_R^2}{M_t^2} +
\frac{1}{12} \log^2 \frac{\mu_R^2}{M_t^2}
\right) \right. \right. \nonumber \\
& & \left. \left. - \frac{4}{3} \sum_{1\leq i\leq N_F} \Delta \left(
\frac{M_i}{M_t} \right) \right]
\left(\frac{\alpha_s^{(N_F)}(\mu_R^2)}{\pi} \right)^2 \right\}
+ {\cal O}(\alpha_s^3)
\end{eqnarray}
where the mass-dependent term involving the light flavours can be
approximated by
\begin{equation}
\Delta(x) = \frac{\pi^2}{8}~x - 0.579~x^2 + 0.230~x^3
\end{equation}
the QCD corrections are formally different from the $\overline{\rm MS}$
case above only for the coefficients $\delta_3, \delta_4$ and $\eta_4$,
\begin{eqnarray}
\delta_3 & = & \frac{897943}{9216} \zeta_3 - \frac{2761331}{41472} +
\frac{209}{64} L_t^2 + \frac{2417}{288} L_t \nonumber \\
& + & N_F \left(\frac{58723}{20736} - \frac{110779}{13824} \zeta_3 +
\frac{23}{32} L_t^2 + \frac{91}{54} L_t \right) + N_F^2
\left(-\frac{L_t^2}{18} + \frac{77}{1728} L_t - \frac{6865}{31104}
\right) \nonumber \\
\delta_4 & = & 
-\frac{121}{1440} N_F \log^{5} 2 +\frac{3751}{2880}
\log^{5} 2+\frac{685}{41472} N_F^2 \log^{4} 2 +\frac{11679301}{
1741824} N_F \log^{4} 2 \nonumber \\
& - & \frac{93970579}{870912}
\log^{4} 2+\frac{121}{144} N_F \zeta_2 \log^{3} 2-\frac{3751}{288}
\zeta_2 \log^{3} 2-\frac{685}{6912} N_F^{2} \zeta_2 \log^{2 } 2
\nonumber \\
& - & \frac{11679301}{290304} N_F \zeta_2 \log^{2} 2 +
\frac{93970579}{145152} \zeta_2 \log^{2} 2 + \frac{4}{9} N_F \zeta_2 \log 2
+\frac{19}{12} \zeta_2 \log 2 \nonumber \\
& + & \frac{2057}{192} N_F \zeta_4
\log 2 - \frac{63767}{384} \zeta_4 \log 2
+ \frac{685}{1728} N_F^{2} a_4
+\frac{11679301}{72576} N_F a_4 -\frac{93970579}{36288}
a_4 \nonumber \\
& + & \frac{121}{12} N_F a_5
- \frac{3751}{24} a_5+\frac{211}{1728} N_F^{3}
\zeta_3
- \frac{270407}{1492992} N_F^{3} - \frac{2}{9} N_F^2 \zeta_2
+\frac{4091305}{331776} N_F^{2}
\zeta_3 \nonumber \\
& - & \frac{576757}{ 55296} N_F^{2} \zeta_4-\frac{115}{384} N_F^{2}
\zeta_5-\frac{161627}{82944} N_F^{2}
- \frac{151369}{725760} N_F
X_0 + \frac{13}{24} N_F \zeta_2 + \frac{19}{4}\zeta_2 \nonumber \\
& - & \frac{12175960469}{38707200} N_F \zeta_3+\frac{608462731}{ 11612160}
N_F \zeta_4+\frac{313489}{6912} N_F
\zeta_5
+ \frac{80863176383}{522547200} N_F \nonumber \\
& + & \frac{4692439}{1451520}
X_0+\frac{28113533041}{19353600} \zeta_3+\frac{4674213853}{ 2903040}
\zeta_4-\frac{807193}{1728} \zeta_5
- \frac{831703495799}{522547200} \nonumber \\
& + & \left(\frac{481}{5184}
N_F^{3}+\frac{28297}{9216} N_F^{2} \zeta_3-\frac{22687}{3456}
N_F^{2}-\frac{32257}{288} N_F \zeta_3
+ \frac{5581849}{41472} N_F+ \frac{9364157}{12288} \zeta_3 \right. \nonumber \\
& & \left. -\frac{46543033}{55296}\right) L_t
+ \left(-\frac{77}{6912}
N_F^{3}-\frac{5107}{13824} N_F^{2}+\frac{12547}{2304} N_F
+\frac{14747}{384}\right) L_t^2 \nonumber \\
& + & \left(\frac{1}{108} N_F^{3}-\frac{157}{576} N_F^{2}+\frac{275}{192}
N_F+\frac{2299}{256} \right) L_t^3 + \frac{4}{3} \left( \frac{2}{3} N_F +
\frac{19}{8}\right) \sum_{1\leq i\leq N_F} \Delta \left( \frac{M_i}{M_t}
\right) \nonumber \\
\eta_4 & = & 
\frac{481}{2592} N_F^{3}+N_F^{2} \left( \frac{28297}{4608}
\zeta_3-\frac{392069}{31104} \right)
+N_F \left(\frac{442189}{1728}-\frac{2985893}{13824} \zeta_3 \right)
\nonumber \\
& + & \frac{26296585}{18432} \zeta_3-
\frac{141262589}{82944}
+ \left(-\frac{77}{1728} N_F^{3}-\frac{2957}{3456}
N_F^{2}+\frac{18241}{1728} N_F+\frac{59195}{576} \right) L_t \nonumber \\
& + & \left(\frac{N_F^{3}}{18} -\frac{455}{288} N_F^{2}+\frac{63}{8}
N_F+\frac{6479}{128}\right) L_t^2
\end{eqnarray}
For the on-shell top-quark mass we obtain numerically for $N_F=5$ light
flavours
\begin{eqnarray}
\delta_3 & = & 11.0154 + 17.9323 L_t + 5.4705 L_t^2 \nonumber \\
\delta_4 & = & 125.7997 + 13.8777 L_t + 55.0041 L_t^2 + 10.4851 L_t^3 +
7.6111 \sum_{1\leq i\leq N_F} \Delta \left( \frac{M_i}{M_t} \right)
\nonumber \\
\eta_4 & = & -114.2461 + 128.5894 L_t + 57.4401 L_t^2
\end{eqnarray}
The explicit expansion of the Lagrangian of Eq.~(\ref{eq:leff}) in powers
of the Higgs field results in
\begin{equation}
{\cal L}_{eff} = \frac{\alpha_s}{12\pi} \left\{ \sum_{n=1}^\infty \Delta_n
\frac{(-1)^{n-1}}{n} \left(\frac{H}{v}\right)^n \right\}
G^{a\mu\nu} G^a_{\mu\nu}
\end{equation}
with the QCD corrections up to N$^4$LO
\begin{eqnarray}
\Delta_1 & = & 1 + \delta_1 \frac{\alpha_s}{\pi} + \delta_2 \left(
\frac{\alpha_s}{\pi} \right)^2 + \delta_3 \left(
\frac{\alpha_s}{\pi} \right)^3 + \delta_4 \left(
\frac{\alpha_s}{\pi} \right)^4 + {\cal O}(\alpha_s^5) \nonumber \\
\Delta_2 & = & 1 + \delta_1 \frac{\alpha_s}{\pi} + (\delta_2+\eta_2) \left(
\frac{\alpha_s}{\pi} \right)^2 + (\delta_3+\eta_3) \left(
\frac{\alpha_s}{\pi} \right)^3 + (\delta_4+\eta_4) \left(
\frac{\alpha_s}{\pi} \right)^4 + {\cal O}(\alpha_s^5) \nonumber \\
\Delta_3 & = & 1 + \delta_1 \frac{\alpha_s}{\pi} + \left(\delta_2+\frac{3}{2}
\eta_2\right) \left(\frac{\alpha_s}{\pi} \right)^2 +
\left(\delta_3+\frac{3}{2} \eta_3+\rho_3\right) \left(
\frac{\alpha_s}{\pi} \right)^3 \nonumber \\
& + & \left(\delta_4+\frac{3}{2}
\eta_4+\rho_4\right) \left( \frac{\alpha_s}{\pi} \right)^4
+ {\cal O}(\alpha_s^5) \nonumber \\
\Delta_4 & = & 1 + \delta_1 \frac{\alpha_s}{\pi} + \left(\delta_2+\frac{11}{6}
\eta_2\right) \left(\frac{\alpha_s}{\pi} \right)^2 +
\left(\delta_3+\frac{11}{6} \eta_3+2 \rho_3\right) \left(
\frac{\alpha_s}{\pi} \right)^3 \nonumber \\
& + & \left(\delta_4+\frac{11}{6} \eta_4+2
\rho_4 + \sigma_4 \right) \left( \frac{\alpha_s}{\pi} \right)^4
+ {\cal O}(\alpha_s^5) \nonumber \\
\Delta_5 & = & 1 + \delta_1 \frac{\alpha_s}{\pi} +
\left(\delta_2+\frac{25}{12}
\eta_2\right) \left(\frac{\alpha_s}{\pi} \right)^2 +
\left(\delta_3+\frac{25}{12} \eta_3+\frac{35}{12} \rho_3\right) \left(
\frac{\alpha_s}{\pi} \right)^3 \nonumber \\
& + & \left(\delta_4+\frac{25}{12} \eta_4+\frac{35}{12} \rho_4 +
\frac{5}{2} \sigma_4 \right) \left( \frac{\alpha_s}{\pi} \right)^4
+ {\cal O}(\alpha_s^5)
\end{eqnarray}
for up to five external Higgs bosons. It should be noted that the
coefficients $\delta_{1-4}$ of the single-Higgs term $\Delta_1$ agree
with previous results up to N$^4$LO \cite{limit,decoup1,decoup,leff},
while the coefficient $\eta_2$ of the double-Higgs contribution
$\Delta_2$ agrees with the explicit diagrammatic calculation of
Ref.~\cite{gghhnnlo2}.

Connecting our approach to derive the effective Lagrangian to the method
of Refs.~\cite{decoup1, decoup} for the single-Higgs case we can easily derive
their final relation,
\begin{equation}
C_H = -\frac{1}{4} \zeta_{\alpha_s}~g_t\partial_{m_t}
\frac{1}{\zeta_{\alpha_s}} = \frac{1}{2v} \frac{m_t^2\partial}{\partial
(m_t^2)} \log \zeta_{\alpha_s}
\end{equation}
with $g_t = m_t/v$, $\partial_{m_t} = \partial/\partial m_t$ and $C_H$
denoting the full coefficient in front of the operator $G^{a\mu\nu}
G^a_{\mu\nu} H$. This expression agrees with Refs.~\cite{decoup1,
decoup}. For the double-Higgs case we arrive at
\begin{equation}
C_{HH} = \frac{1}{8} \zeta_{\alpha_s}~g_t^2\partial_{m_t}^2
\frac{1}{\zeta_{\alpha_s}} = \frac{1}{4v^2}\left\{ \left(
\frac{m_t\partial_{m_t} \zeta_{\alpha_s}}{\zeta_{\alpha_s}} \right)^2
-\frac{m_t^2\partial_{m_t}^2\zeta_{\alpha_s}}{2\zeta_{\alpha_s}}
\right\}
\end{equation}
where $C_{HH}$ denotes the coefficient in front of the operator
$G^{a\mu\nu} G^a_{\mu\nu} H^2$.

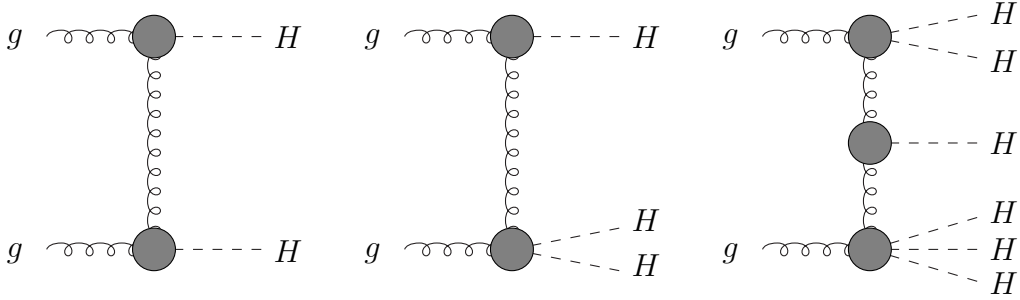
\begin{figure}[hbt]
\SetScale{0.8}
\begin{picture}(100,90)(-50,0)
\Gluon(0,100)(50,100){3}{4}
\Gluon(0,0)(50,0){3}{4}
\Gluon(50,0)(50,100){3}{10}
\DashLine(50,100)(100,100){5}
\DashLine(50,0)(100,0){5}
\GCirc(50,100){10}{0.5}
\GCirc(50,0){10}{0.5}
\put(85,76){$H$}
\put(85,-5){$H$}
\put(-15,-3){$g$}
\put(-15,78){$g$}
\SetScale{1}
\end{picture}
\begin{picture}(100,90)(-80,0)
\Gluon(0,100)(50,100){3}{4}
\Gluon(0,0)(50,0){3}{4}
\Gluon(50,0)(50,100){3}{10}
\DashLine(50,100)(100,100){5}
\DashLine(50,0)(100,10){5}
\DashLine(50,0)(100,-10){5}
\GCirc(50,100){10}{0.5}
\GCirc(50,0){10}{0.5}
\put(85,76){$H$}
\put(85,7){$H$}
\put(85,-10){$H$}
\put(-15,-3){$g$}
\put(-15,78){$g$}
\SetScale{1}
\end{picture}
\begin{picture}(100,90)(-110,0)
\Gluon(0,100)(50,100){3}{4}
\Gluon(0,0)(50,0){3}{4}
\Gluon(50,0)(50,100){3}{10}
\DashLine(50,100)(100,110){5}
\DashLine(50,100)(100,90){5}
\DashLine(50,50)(100,50){5}
\DashLine(50,0)(100,15){5}
\DashLine(50,0)(100,0){5}
\DashLine(50,0)(100,-15){5}
\GCirc(50,100){10}{0.5}
\GCirc(50,50){10}{0.5}
\GCirc(50,0){10}{0.5}
\put(85,85){$H$}
\put(85,67){$H$}
\put(85,36){$H$}
\put(85,10){$H$}
\put(85,-4){$H$}
\put(85,-17){$H$}
\put(-15,-3){$g$}
\put(-15,78){$g$}
\SetScale{1}
\end{picture} \\
\caption[]{\it \label{fg:1pr} Typical one-particle-reducible Feynman
diagrams for multi-Higgs boson production.}
\end{figure}

A final comment addresses the removal of one-particle-reducible
contributions in Eq.~(\ref{eq:pit}): this corresponds to the removal of
one-particle-reducible diagrams of the type shown in Fig.~\ref{fg:1pr}
after attaching external Higgs bosons according to Eq.~(\ref{eq:shift}).
We have checked this correspondence explicitly for Higgs boson pair
production in the heavy-top-quark limit at NLO \cite{gghhnlo}.

\section{Several Heavy Quarks}
%        ====================
Starting from the expression of the effective single-Higgs coupling to
gluons of Ref.~\cite{leffnh} with $N_H$ heavy quarks contributing we can
reconstruct the corresponding logarithmic parts of the function $\Pi_Q$,
\begin{eqnarray}
{\cal L}_{g} & = & -\frac{1-\Pi_Q}{4} \hat G^{a\mu\nu} \hat G^a_{\mu\nu}
\nonumber \\
\Pi_Q & = & \sum_n C_n \left(\frac{\alpha_s^{(N_F+N_H)}}{\pi}\right)^n
\end{eqnarray}
with the perturbative coefficients up to third order
\begin{eqnarray}
C_1 & = & -\frac{N_H}{6} L_Q \\
C_2 & = & N_H \left[ \frac{11}{72} - \frac{11}{24} L_Q \right] \nonumber \\
C_3 & = & C_{30} - N_H \left( \frac{1877}{1152} - \frac{77}{3456} N_H -
\frac{67}{576} N_F \right)  L_Q - N_H \left( \frac{19}{192}
- \frac{11}{288} N_H + \frac{N_F}{36} \right) L_Q^2 \nonumber
\end{eqnarray}
where $\hat G^{a\mu\nu}$ denotes the gluonic field-strength operator of
colour-SU(3) in the low-energy limit with $N_F+N_H$ active flavours.
The logarithm is defined as
\begin{equation}
L_Q = \frac{1}{N_H} \sum_{i=1}^{N_H} \log \left(\frac{\mu_R^2}{M_i^2}
\right)
\end{equation}
For the derivation of the effective Lagrangian for the gluonic Higgs
coupling the constant $C_{30}$ is irrelevant. Performing the
replacement\footnote{Here we assume SM-type couplings of the heavy
quarks to the Higgs boson as e.g.~for a sequential 4th fermion
generation. For the case of different couplings and $N_S$ scalar Higgs
bosons this shift has to be replaced by $\log(1+H/v) \to
\sum_{i=1}^{N_H} \log \left(1 + \sum_{j=1}^{N_S} c_{ij} H_j/v \right) /
N_H$ in all subsequent steps, where the factors $c_{ij}$ denote the
Higgs Yukawa couplings normalized to the SM-Higgs coupling.}
\begin{equation}
L_Q \to \bar L_Q = L_Q - 2\log\left(1 + \frac{H}{v} \right) \qquad
\mbox{and} \qquad
\Pi_Q \to \bar \Pi_Q
\end{equation}
and decoupling the heavy quarks from the strong coupling constant
$\alpha_s$ by
\begin{eqnarray}
\alpha_s^{(N_F+N_H)}(\mu_R^2) & = & \alpha_s^{(N_F)}(\mu_R^2) \left\{ 1 +
\frac{\alpha_s^{(N_F)}(\mu_R^2)}{\pi} N_H \frac{L_Q}{6} \right.
\nonumber \\
& + & \left. \left( \frac{\alpha_s^{(N_F)}(\mu_R^2)}{\pi} \right)^2 N_H
\left[ -\frac{11}{72} + \frac{11}{24} L_Q + N_H \frac{L_Q^2}{36}
\right] \right\} + {\cal O}(\alpha_s^4)
\end{eqnarray}
and from the gluon-field-strength operator we arrive at the effective
Lagrangian for the gluonic Higgs couplings up to NNLO
\begin{equation}
{\cal L}_{eff} = N_H \frac{\alpha_s}{12\pi} \left\{ (1 + \delta) \log
\left(1+\frac{H}{v}\right) - \frac{\eta}{2} \log^2
\left(1+\frac{H}{v}\right) \right\} G^{a\mu\nu} G^a_{\mu\nu}
\label{eq:leffnh}
\end{equation}
with the QCD corrections up to NNLO
\begin{eqnarray}
\delta & = & \delta_1 \frac{\alpha_s}{\pi} + \delta_2 \left(
\frac{\alpha_s}{\pi} \right)^2 + {\cal O}(\alpha_s^3) \nonumber \\
\eta & = & \eta_2 \left( \frac{\alpha_s}{\pi} \right)^2 + {\cal O}(\alpha_s^3)
\end{eqnarray}
The explicit perturbative coefficients read
\begin{eqnarray}
\delta_1 & = & \frac{11}{4} \nonumber \\
\delta_2 & = & \frac{1877}{192} - \frac{77}{576} N_H + \frac{19}{16} L_Q + N_F
\left(\frac{L_Q}{3}-\frac{67}{96} \right) \nonumber \\
\eta_2 & = & \frac{19}{8} - \frac{11}{12} N_H + \frac{2}{3} N_F
\end{eqnarray}
The result for $\delta_2$ in the single-Higgs case agrees with the
results of Refs.~\cite{leffnh,leffnh0}. The NNLO results for more than
one external Higgs boson are new.

\section{Conclusions}
%        ===========
In this work we have derived effective (multi-)Higgs couplings to gluons
after integrating out all heavy quarks mediating these couplings. The
effective Lagrangians can be used for the computation of the production
of one or several Higgs bosons in gluon fusion at hadron colliders in
the limit of heavy quarks. In the SM we have extended the effective
Lagrangian for double-Higgs couplings to gluons to N$^4$LO and derived
for the first time the N$^4$LO Lagrangian for more than two SM Higgs
bosons. In the second part we extended the analysis to the case of
several heavy quarks coupling to the Higgs bosons up to NNLO. We
reproduced the existing NNLO results for the single-Higgs case.  We have
derived these effective Lagrangians from their connection to the
decoupling relations of the strong coupling constant.

\section*{Acknowledgments}
%         ===============
We are grateful to M.~M\"uhlleitner and A.~Signer for carefully reading
the manuscript and useful comments. This work is supported in part by
the Research Executive Agency (REA) of the European Union under the
Grant No.~PITN-GA-2012-316704 (Higgstools).

\end{document}